\documentclass[runningheads]{svmult}

\usepackage{makeidx}   
\usepackage{graphicx}  
\usepackage{subeqnar}  
\usepackage{multicol}  
\usepackage{physprbb}  
\makeindex             

\usepackage{psfig}

\begin{document}
\title*{Multi-Conjugate Adaptive Optics
\protect\newline with Laser Guide Stars}
\toctitle{Multi-Conjugate Adaptive Optics
\protect\newline with Laser Guide Stars}
%
%
\titlerunning{MCAO with LGS}
%
\author{R.I. Davies\inst{1}
\and D. Bonaccini\inst{2}
\and S. Rabien\inst{1}
\and W. Hackenberg\inst{2}
\and T. Ott\inst{1}
\and S. Hippler\inst{3}
\and U. Neumann\inst{3}
\and M. Barden\inst{1}
\and M. Lehnert\inst{1}
\and F. Eisenhauer\inst{1}
\and R. Genzel\inst{1}}
\authorrunning{R.I. Davies et al.}
%
%
\institute{Max-Planck-Institut f\"ur extraterrestrische Physik,
85741 Garching, Germany
\and European Southern Observatory, 
85748 Garching, Germany
\and Max-Planck-Institut f\"ur Astronomie, 
69117 Heidelberg, Germany}

\maketitle              

\begin{abstract}
The Laser Guide Star Facility for the VLT will be commissioned
during 2003. 
This, of course, can only be the first step towards enhancing the
scientific output of the VLT through adaptive optics (AO). 
So in this contribution we propose the development of a 
laser guide star (LGS) 
multi-conjugate adaptive optics (MCAO) system. 
We consider geometries for the adaptive optics system, and discuss the
technology needed to project up to 5 laser beacons.
The rationale for such a project is provided by an outline of two primary
science drivers. 
\end{abstract}

\section{Introduction}

The progress that has been made in understanding advanced adaptive
optics techniques in recent years now allows us to design systems that are
tailored to particular astronomical requirements.
For a limited amount of light from one or more reference sources, this
is effectively a trade off between finely sampling (spatially and
temporally) the atmospheric
turbulence over a small area to obtain extremely high Strehl ratios over a
limited field of view, or broadly sampling the turbulence over a large
area to obtain moderate performance over a wide field.

Here we discuss which characteristics might be desirable in a system
that would have the widest possible applications for 1--2.5\,$\mu$m
observations.
We argue that laser guide stars are mandatory if these are to be
achieved. 
The particular science drivers we consider -- observational
cosmologogy and the initial mass function (IMF) in massive galactic star
clusters -- indicate the potential wealth which can be reaped with LGS
MCAO.

\section{MCAO characteristics}

What an astronomer really wants from AO, and which in
fact can only be delivered by a MCAO system, can be summarised
simply as:
{\it obtaining, with high sky coverage, a uniform high Strehl over a
large field of view, that field being the same at all wavelengths}.
Considering each of the points in turn:

\begin{itemize}
\item
{\bf High Strehl ratio} 
is the raison d'\^etre for AO systems, to provide better
sensitivity and resolution. The problem of limited Strehl with a
single LGS (the cone effect) is solved with MCAO.
Strehl ratios greater than 50\% in the K-band, or equivalent in other
bands, should be achievable.
\item
{\bf Uniform Strehl ratio over a large field of view}
($\sim$1$^\prime$ can be corrected with relatively few guide
stars) is the primary reason for using MCAO, solving
the isoplanatic limitation of normal AO.
It assures
reliable photometry (eg for star clusters or extended objects) and 
uniform sensitivity (eg for number counts); and a PSF reference would
usually be found in the field.
\item
{\bf The same field of view at all wavelengths}
comes by default with MCAO: the field of view attained is
due to the height of the atmospheric turbulence rather than
the $\lambda$-dependent isoplanatic angle.
For multi-colour studies, properties of the same objects can
be measured in different bands.
\item
{\bf High sky coverage}
ensures that as many programmes as possible can be executed, and
is necessary if MCAO is to have an impact on astrophysics.
\end{itemize}

\section{Natural Guide Stars}

The preferred option for MCAO in terms of cost and complexity would
certainly be to use NGS if the requirements above could be fulfilled.
We briefly consider the implications of this on the two designs for
MCAO that have been proposed.

\subsection{Classical MCAO}

The original MCAO concept involves sensing multiple guide stars 
separately to provide measurements of the total wavefront
aberrations in several directions.
By making some assumptions (or additional measurements) about the
height and number of turbulent layers, it 
is possible to derive the phase shifts induced by each.
It is then possible to correct these with a DM conjugated to each layer.

The advantage of this method is that relatively few guide stars are
needed, typically 3--5 to correct $\sim$1$^\prime$.
However, only the area between the stars is corrected well, and then only if
they are less than 30--60$^{\prime\prime}$ apart.
Furthermore, the maximum
Strehl ratio towards any given guide star depends on its magnitude.
A uniform high Strehl can therefore be achieved for only a very
few sources.
In the vast majority of other possible targets -- which is still
limited -- the Strehl ratio will vary considerably across the field.
As a consequence of compromising the performance in this way,
the resolution and sensitivity will vary greatly and the data
analysis become extremely difficult, losing the advantage that MCAO
should have over standard AO.

\subsection{Layer oriented MCAO}

An alternative concept (Ragazzoni et al. 2000) avoids these problems
to some extent by using the light from many stars in the field rather
than just the few brighter ones.
The design uses pyramid wavefront sensors, which allow the light
from all the sources to be combined before being sensed.
Each of 2--3 turbulent layers is corrected independently and requires
one DM and one detector, both of which are conjugated to it.
This vastly simplifies the computations, but if many stars are used
the optics in the AO system becomes rather more complex.
Additionally, the light from all the stars in a field is
typically only 14--15\,mag, rather less than that needed for optimal
correction (eg NAOS requires stars of $\sim$12\,mag for best
performance, Rousset et al. 2000).
If on the other hand fewer stars are used, then the problems of
variable Strehl across the field return due to strong variations in
the signal-to-noise with position.

\section{Laser Guide Stars}

One solution to the problems outlined above is to use artifical guide
stars.
This applies to either of the concepts mentioned,
or even new ones -- indeed, one scheme optimised for use with LGS is
currently under study. 
Sodium laser guide stars (and also Rayleigh beacons, although these
are not considered here because there are additional MCAO
design issues which require further development)
allow one to have regularly placed bright sources of the same magnitude.
The power and positioning can be adjusted to produce the required
uniform high strehl (Berkefeld et al 2001).
Simulations by Rigaut et al. (2000) for the Gemini LGS-MCAO system
indicate that with 5 LGSs the Strehl ratio varies by only a few percent
from the peak value of 60--70\% (at K) over a field of
1--2$^\prime$ at zenith.

The VLT Laser Guide Star Facility (LGSF) Launch Telescope is already
designed to cope with 5 lasers, 
and is diffraction limited over a radius of $>$1$^\prime$.
The pointing of each beam is controlled by a piezo system, allowing
them to be moved in a centred square geometry anywhere within this
region.

A number of objections to sodium LGS have been raised.
We show below that these are not a major hindrance to the
implementation of LGS MCAO.
\newline
1) The first generation of LGS on 3--4-m telescopes were difficult and
inefficient to use, but experience with such
systems as ALFA (Hippler et al. 2000, Davies et al. 2000) will
lead to significant improvements for LGS at 8--10-m 
telescopes.
\newline
2) Up to 3 NGS are needed to sense the low order modes (the
inability to sense tip-tilt from a LGS projects into astigmatism and
focus for MCAO).
However, the requirements on magnitude and position are far less
stringent than before and do not impose strong constraints on sky
coverage.
\newline
3) The LGS elongation is less than
1$^{\prime\prime}$ as long as the launch telescope is behind the
VLT secondary.
It depends primarily on the horizontal separation
of the launch and detection points, which need only be 4-m
for an 8-m telescope.
\newline
4) Light scattered in the Rayleigh cone interferes with the wavefront
sensing.
A simple but wasteful option is to pulse a continuous wave laser via 
extra-cavity modulation (eg as was done on much faster timescales
for LIDAR on ALFA, Butler et al. 2000).
Alternatively, careful baffling of the Rayleigh light at its focal
plane in the AO system may alleviate much of the problem without
greatly compromising the brightness or shape of the LGS.

\subsection{Laser Power}

In order for a LGS to provide 10$^6$\,ph\,m$^{-2}$\,s$^{-1}$
(equivalent) at the Nasmyth focus even in `poor' conditions, a laser
power of 10\,W is needed.
This scales to 50\,W for MCAO with 5 lasers.
As a result ESO is currently pursuing a research programme to develop
589\,nm continuous wave fibre lasers, each producing 10\,W with a
500\,MHz bandwidth matched to the mesospheric sodium absorption profile.
These systems would be relatively simple to operate and require no
additional beam relay since the fibre laser itself could extend to the
launch telescope.

A short-term alternative would be to upgrade PARSEC, the laser which
will initially be used in the LGSF.
PARSEC is a ring dye laser system: pump lasers emitting green
532\,nm light excite molecules in a high speed dye jet, which
decay to produce the 589\,nm sodium line emission.
It has a master laser operating at low power which can be locked to
the right frequency and tuned to produce a high quality output
beam.
This seeds an amplifier which provides a scalable boost to the
final output power.
Although a full upgrade 
could in principle provide 50\,W,
a very simple change that involves only pumping the existing dye
jets with more power, could increase the 589\,nm output to around 25\,W.

In this case each LGS would be $\sim$0.8\,mag fainter.
Such a loss does not compromise AO performance much and, during
most observing, would be recovered:
\newline
1) The 10\,W specification for each LGS takes into account observing at
an airmass of 2; at an airmass of 1.5, where the seeing is anyway
better, each LGS would be 0.3\,mag brighter.
\newline
2) A sodium column density above 3$\times$10$^{13}$\,m$^{-2}$ (which
occurs more than 60\% of the time, Ge et al. 1998) rather than the
minimum specified
2$\times$10$^{13}$\,m$^{-2}$ (which occurs 80\% of the time), would
gain another 0.45\,mag.
\newline
3) At least 0.1\,mag is expected to be gained through advances
in the fibre beam relay system for PARSEC.

\section{Science Cases}

\subsection{Observational Cosmology}

With current AO systems, the only way to proceed with
observational cosmology is to study objects close to bright stars:
within $\sim$30$^{\prime\prime}$ for the K-band, or
$\sim$15$^{\prime\prime}$ in the J-band (to detect H$\alpha$ in
objects at $z$$\sim$1).
The other is to use MCAO.

\begin{figure}[h]
\begin{minipage}[m]{5.5cm}
\centerline{\psfig{file=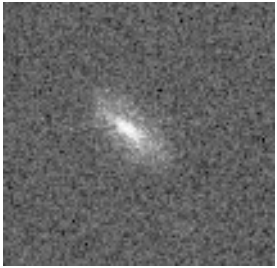,width=3.6cm}}
\end{minipage}
\begin{minipage}[m]{6cm}
\caption{5$^{\prime\prime}$$\times$5$^{\prime\prime}$ simulation of a
1\,hr integration on the VLT with a 50\% Strehl showing a K=19 galaxy,
in this case NGC\,253 redshifted to $z$$\sim$1. 
The fainter extended emission is clearly seen, allowing a
morphological classification.\label{fig:ngc253}}
\end{minipage}
\end{figure}

Observations with ALFA of the 1$^\prime$ field around such a star,
with 30\% K-band Strehl, have revealed a surprisingly high
number of galaxies and faint point sources at K$\sim$19--20.5 (Davies
et al. 2001), at which magnitude the mean redshift is
$z$=0.7 (Cowie et al. 1996).
The resolution of 0.15$^{\prime\prime}$, already better than that
attained with NICMOS on the HST, is sufficient to easily measure the
light profiles, one of the robust parameters used for the
classification of high redshift objects 
in studies of the cosmological evolution of galaxies 
(Abraham 1999).
Simulations indicate that a 1\,hour integration on the VLT with 50\%
Strehl in the K-band is
sufficient to see the more detailed morphology of K$\sim$19 galaxies,
as shown in Fig.~\ref{fig:ngc253}.
As a result of the difference between observed and rest frames, 
such data would be far more successful at classification of high
redshift objects than the similar resolution HST I-band data, for
which at $z$=0.9 about 24\% of spirals are misclassified as
peculiars (Brinchmann et al. 1998).

\begin{figure}
\centerline{\psfig{file=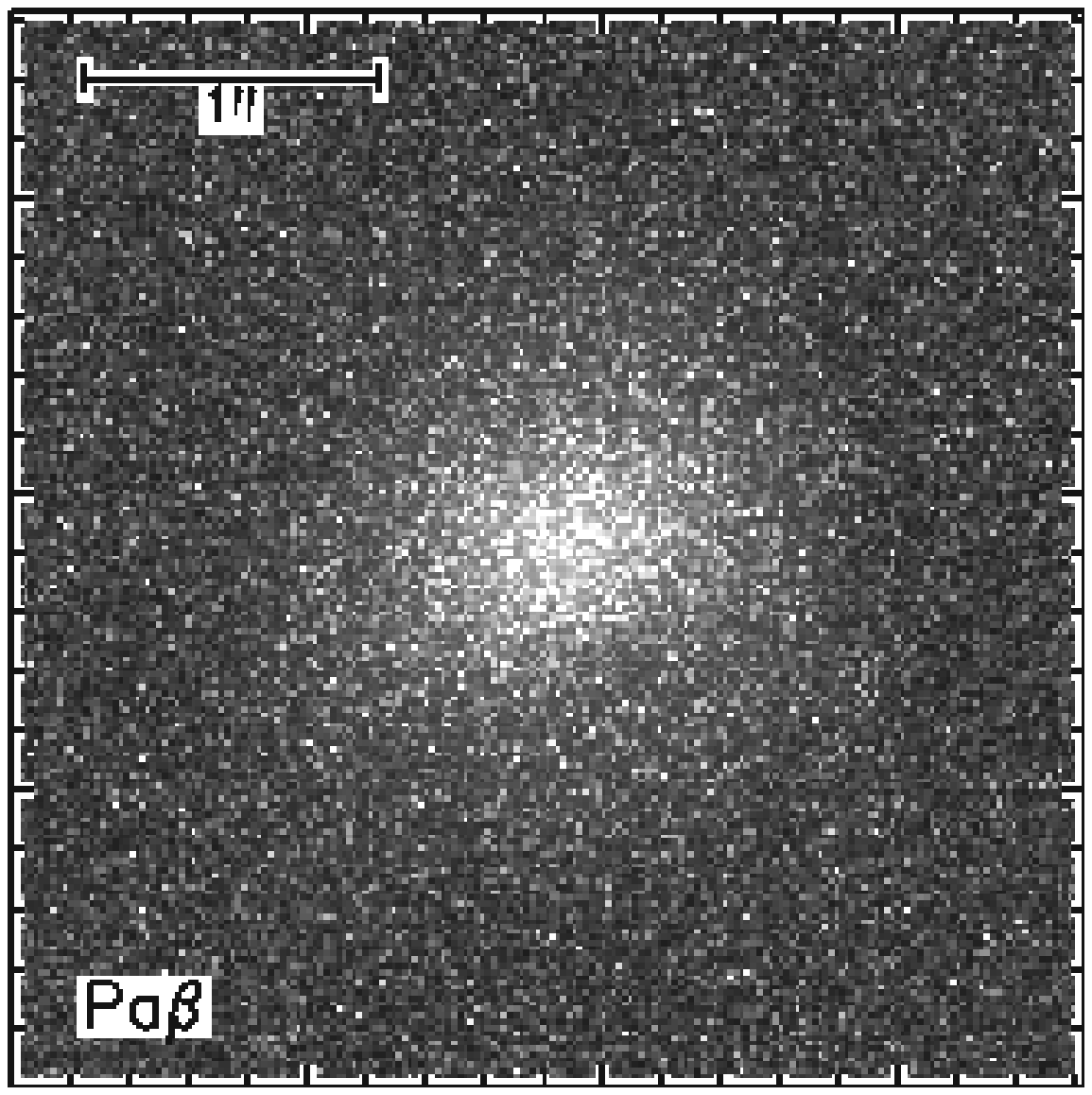,width=3.6cm}\hspace{2mm}\psfig{file=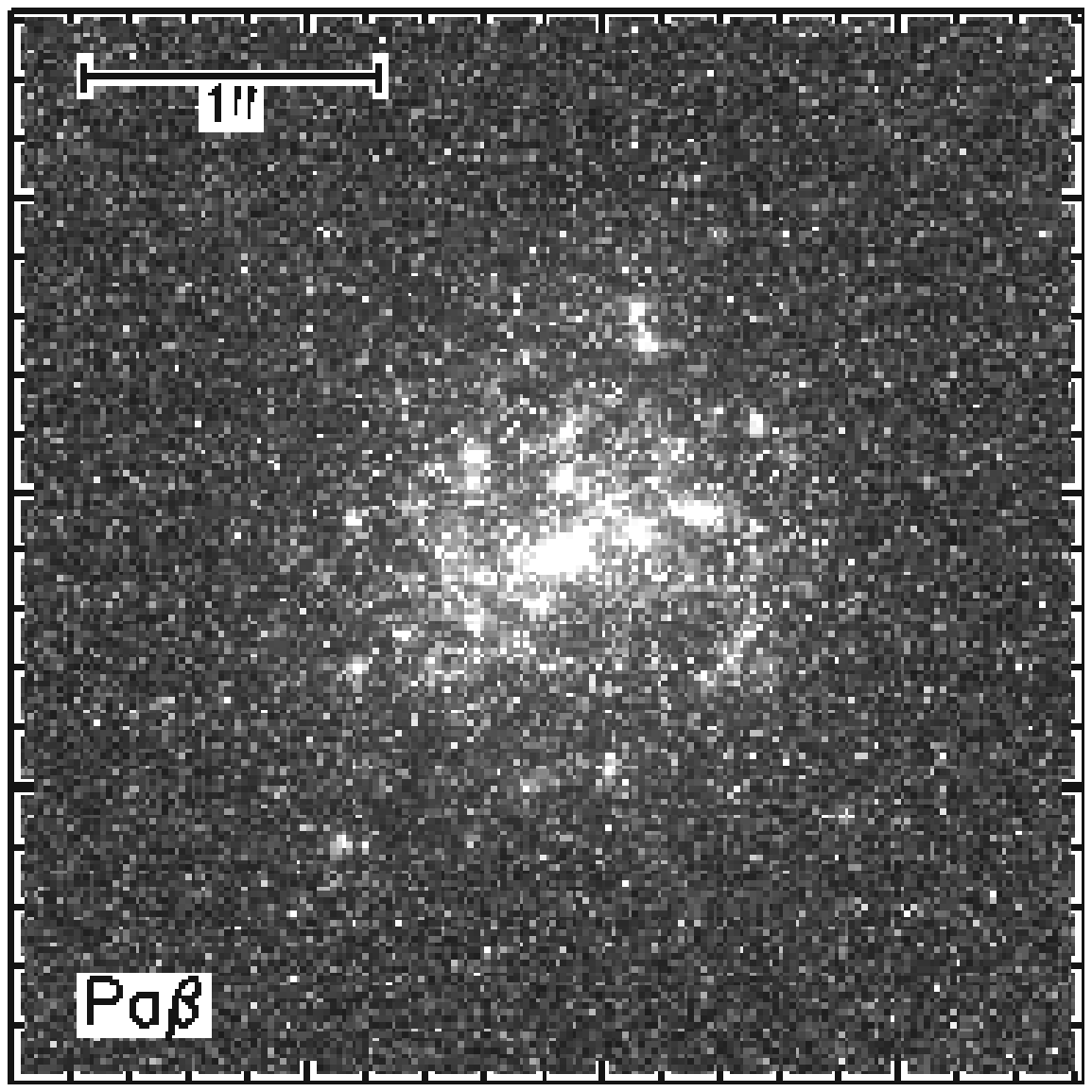,width=3.6cm}}
\caption{H$\alpha$ line map of IC\,342 as it would appear
at $z$=0.7 with the Pa$\beta$ line imaged in the K-band.
Left is seeing limited; 
right is with a 50\% Strehl, showing that the H{\sc ii} regions are
clustered on scales matched to the resolution of the VLT.
\label{fig:ic342}}
\end{figure}

Another aspect of high redshift work is the Tully-Fisher
relation, which correlates the rotational velocity of a galaxy's disk
with its absolute magnitude.
Changes in the relation at higher redshifts (eg Barden \& Lehnert,
priv. comm.) clearly indicate evolution
in the mass to light ratio of disks, but whether this is purely
passive evolution or indicates something more requires careful analysis
of galaxy evolution models.
This type of work is extremely difficult with ground-based telescopes:
the galaxies are typically 1$^{\prime\prime}$ across, and so even at
good observing sites it is
not only difficult to align a slit along the major axis but the
rotation curves are barely resolved.
AO would have 2 benefits.
The first is to reduce the blending in the rotation curve by putting
the light back where it should be, without necessarily increasing the
surface brightness.
The second is effectively to increase the surface brightness, because the
(clusters of) H{\sc ii} regions responsible for the emission lines
appear as point sources at the diffraction limit of the VLT as shown
in Fig.~\ref{fig:ic342}.
For such work, it is undesirable and unnecessary to fully sample the
PSF:
the former because the sensitivity would be limited by read noise and
dark current, the latter because only a few independent points are
needed to measure the rotation curve.
Infact, pixels of $\sim$0.075$^{\prime\prime}$ are ideal, equivalent
to $\sim$500\,pc at $z$=1.
The pixel gain (ie increase in flux that should be in a pixel, compared to
the seeing limited case) is $\times$25 for 50\% Strehl in
the K-band or,
for a similar AO performance, $\times$15 for 18\% Strehl in the
J-band.
Using MCAO to extend the current work to lower surface brightness
galaxies or to higher redshifts, and to measure several galaxies
simultaneously (with a multi-object spectrograph or integral field
unit), would enable important progress in understanding galaxy evolution.

\subsection{Massive Star Clusters}

A rather different application for MCAO is in the study of the IMF in
massive star clusters.
Using ADONIS, Eisenhauer et al. (1998) reached about 20\% K-band Strehl
ratio in the centre of NGC\,3606 and were able to measure its IMF from
about 800 stars -- but only after careful
deconvolution using PSFs appropriate to different parts of the
cluster.
They found that while the IMF above $\sim$30\,M$_\odot$ was similar to
both the Scalo field star IMF and that measured in the Upper Scorpius
association, for stars below 
$\sim$30\,M$_\odot$ the IMF had a much shallower slope.
Later work using the VLT (Brandl et al. 1999) was far more sensitive,
but because of the seeing limited resolution of 0.3--0.4$^{\prime\prime}$,
the uncertainties in correcting for completeness
remain too great to permit a final sensus of the low mass stellar
population -- something that would be possible with MCAO, which could
image the central 2--3\,pc of the cluster in all of the JHK bands and
reach a high Strehl ratio in each.

\section{Conclusion}

We have argued that laser guide stars are essential for MCAO to be
scientifically useful to the astronomical community:
they provide the only way to achieve a high uniform Strehl over a
large field with good sky coverage.

The technical problems in providing sufficient laser power are being
addressed. 
A simple upgrade to PARSEC would provide enough power at least for
most observing conditions, and is a realistic option within the next
few years.

The two science cases represent areas where MCAO could make important
contributions, but which are unattainable without the use of laser
guide stars.

%

\end{document}